# Identification of a *DAGLB* mutation in a Non-Chinese patient with Parkinson's disease.

Christelle Tesson[1], PhD, Mohamed Sofiane Bouchetara[2], MD, Mélanie Ferrien[1], MSc, Suzanne Lesage[1], PhD, Alexis Brice[1,3], MD

[1] Sorbonne Université, Institut du Cerveau - Paris Brain Institute - ICM, Inserm, CNRS, Paris, France

[2] Service de Neurologie, Etablissement hospitalo-universitaire, Oran, Algérie

[3] Assistance Publique Hôpitaux de Paris, Hôpital Pitié-Salpêtrière, Département de Neurologie, Centre d'Investigation Clinique Neurosciences, DMU Neuroscience, Paris, France

Correspondence to: Prof. Alexis Brice, MD; Institut du Cerveau, ICM, Hôpital de la Salpêtrière, Paris, France
E-mail: alexis.brice@icm-institute.org | Tel.: +33 1 57 27 42 63

Liu and colleagues[1] recently reported that biallelic mutations in *DAGLB* are responsible for autosomal recessive early-onset Parkinson's disease (AR-EO-PD). They first identified a homozygous *DAGLB* splice site mutation in a consanguineous family with two affected siblings by combining homozygosity mapping and whole exome sequencing (WES). Through exome data mining in a large cohort of 1,741 unrelated PD cases, they identified three additional mutated index cases. All six patients were of Chinese origin and presented typical EO-PD (≤40 years) with a good response to levodopa. No additional cases outside China have been reported so far.

To assess the causal role of pathogenic variants in *DAGLB* in our French cohort, we used data mining in the exomes of 55 families with AR-PD (n=88 cases), and 629 isolated EO-PD cases (Supplementary Data and Material). We identified a homozygous p.Pro357Leu missense variant in a single consanguineous PD case (Figure 1a). This mutation predicted to be deleterious by 13 out of 18 prediction softwares tested (Supplementary Table 1) affects a conserved amino acid localized in the catalytic domain of the protein, near the pathological p.Asp363Gly mutation described in the previous paper[1] (Figure 1b). It is predicted to destabilize the protein structure according to Dynamut2[2] (Figure 1c and 1d). All rare variants in the homozygous state identified in WES analysis of this patient are available in Supplementary Table 2.

The patient of Algerian origin was born to healthy first-cousins parents and reported no family history of PD. His family and social cirumstances led him to stop his schooling at the age of 15. He presented at 30 years with bradykinesia and later developed postural instability and an akineto-rigid form of the disease, predominant on the right side, with an UPDRS III motor score of 25 and a Hoehn & Yahr score of 3 on "off" stage and of 2 under medication. Following 13 years of disease evolution, his akineto-rigid syndrome worsened with the presence of dysarthria, an UPDRS III score of 78 on "off" stage, and of 45 under medication and a Hoehn & Yahr stage of 4 on the "off" stage which improved to 3 on medication. The patient did not show any frank cognitive impairment or hallucinations. At the last examination, he developed non-motor signs and symptoms, such as depression, constipation, urinary and sleep disturbances. Levodopa treatment led to significant improvement in clinical signs. The patient developed subsequently end-of-dose dyskinesias, motor fluctuations as well as dystonia in the "off" medication state. Brain MRI was normal (see Supplementary Table 3).

Thus, we identified a patient carrying the *DAGLB* p.Pro357Leu mutation localized nearby the previously published p.Asp363Gly[1] variant, reinforcing the fact that *DAGLB* is involved in EO-AR-PD. As the most frequent genes involved in AR-PD (*PRKN*[3], *PINK1*[4]), the *DAGLB*-associated disease presents and evolves like typical PD. Since we screened a population of PD cases with either AR inheritance or EO-PD (< 50 years) and found a single patient among 683 index cases, we conclude that *DAGLB* is a very rare cause of AR-EO-PD. However, we demonstrate that mutations in *DAGLB* are not limited to the Chinese PD population but can also account for PD in North Africa.

**Acknowledgment**

We thank the patients and their families. We thank the DNA and cell Bank .of the ICM for sample preparation. Part of this work was carried out on the iGenSeq and DAC core facilities of the ICM. We thank Dr. Poornima Menon for proofreading the text.

**Authors' Roles**

1) Research project: A. Conception, B. Organization, C. Execution;

2) Statistical Analysis: A. Design, B. Execution, C. Review and Critique;

3) Manuscript: A. Writing of the first draft, B. Review and Critique;

4) Clinical investigation.

Christelle Tesson : 1A, 1B, 1C, 3A

Mohamed Sofiane Bouchetara : 3B, 4

Suzanne Lesage: 1A, 1B, 3B

Alexis Brice: 1A, 1B, 3B, 4

All authors have read and agreed to the published version of the manuscript.

**Funding Sources and Conflict of Interest:**

Part of this work was funded by grants from France Parkinson Association, Fondation de France (n° 00076353), la Fédération pour la Recherche sur le Cerveau (FRC), the program "Investissements d'avenir" (ANR-10-IAIHU-06). Authors report no conflicts of interest related to the paper

**Financial Disclosures for the previous 12 months:**

Christelle Tesson was employed by Paris Brain Institute (ICM) during this work. Suzanne Lesage has received grants from Fondation de la Recherche Médicale (FRM, MND202004011718). Alexis Brice has received grants from Fondation Roger de Spoelberch and Greater Paris University Hospitals (APHP).

**Ethical Compliance Statement**

Informed consent was obtained from all participants, and genetic studies were approved by local ethics committees (INSERM, CCPPRB du Groupe Hospitalier Pitié-Salpêtrière, Paris, France, N° 44814). We confirm that we have read the Journal's position on issues involved in ethical publication and affirm that this work is consistent with those guidelines.

**References**

1. Liu Z, Yang N, Dong J, et al. Deficiency in endocannabinoid synthase DAGLB contributes to early onset Parkinsonism and murine nigral dopaminergic neuron dysfunction. *Nat Commun*. 2022;13:3490. doi:10.1038/s41467-022-31168-9

2. Rodrigues CHM, Pires DEV, Ascher DB. DynaMut2: Assessing changes in stability and flexibility upon single and multiple point missense mutations. *Protein Sci*. 2021;30(1):60-69. doi:10.1002/pro.3942

3. Lücking CB, Dürr A, Bonifati V, et al. Association between Early-Onset Parkinson's Disease and Mutations in the Parkin Gene. *New England Journal of Medicine*. 2000;342(21):1560-1567. doi:10.1056/NEJM200005253422103

4. Ibáñez P, Lesage S, Lohmann E, et al. Mutational analysis of the PINK1 gene in early-onset parkinsonism in Europe and North Africa. *Brain*. 2006;129(Pt 3):686-694. doi:10.1093/brain/awl005

**Figure**

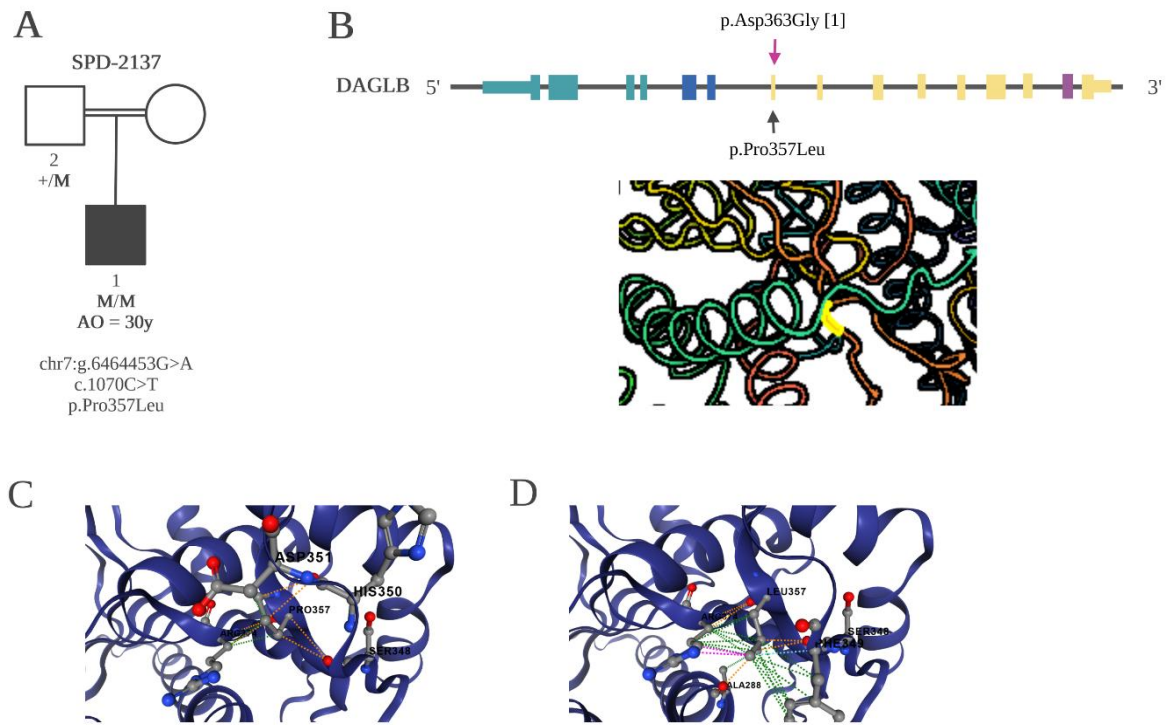

**Figure legends**

**Fig. 1** Segregation and localization of the *DAGLB* p.Pro357Leu mutation. **a** Segregation of the *DAGLB* mutation in the SPD-2137 family. The patient is homozygous for the p.Pro357Leu mutation and his unaffected father is heterozygous for this mutation. **b** Localisation of the *DAGLB* mutation identified in our study nearby the p.Asp363Gly mutation identified in the lead article [1], adapted from [1]. The catalytic domain is indicated in yellow. **c** Prediction of the wild-type DAGLB structure and amino acid interaction, obtained with DynaMut2 [2]. **d** Prediction of the mutated DAGLB structure due to the p.Pro357Leu mutation and amino acid interactions, obtained with DynaMut2 [2]. Predicted interactions are greatly modified compared to the wild type DAGLB ($\Delta\Delta G^{Stability}$ = -0.49kcal/mol) . AO: age at onset, +: Wild type, M: mutated.

Created with BioRender.com

**Supplementary Material and Methods**

**Cohort description**

The cohort assembled by the French and Mediterranean Parkinson's Disease Genetics Study group (FMPD cohort) consisted of 683 index PD patients with available whole-exome sequencing (WES) data. Most of these patients present with early-onset PD. Mutations in the Parkinsonism-associated genes (Supplementary Table 4), were previously excluded using WES and multiplex ligation-dependent probe amplification[1,2]. Expansions in SCA2 and SCA17 were excluded using ExpansionHunter[3,4].

This cohort included 716 patients with PD, 273 females and 443 males, from 683 families. 472 were from Europe, 102 from North Africa, 7 from South Africa, 1 from Asia and 134 of unknown origin. Among them, 55 families with at least two affected siblings (n = 88 patients) were compatible with AR PD, including 13 families (n = 20 patients) with parental consanguinity. 628 patients were isolated cases, including 61 with consanguinity. WES data were available for all index cases, 33 additional affected members and 53 unaffected relatives. The mean age at onset of motor signs was 35.7 years (range: 6 to 77 years).

Informed consent was obtained from all participants, and the genetic studies were approved by local ethics committees (INSERM, CCPPRB du Groupe Hospitalier Pitié-Salpêtrière, Paris, France).

**Exome sequencing and analyses**

WES in the FMPD cohort was performed at the ICM IGenSeq core facility, Integragen service (Evry, France) or through the NIH facility. Exons were captured using the SureSelect Human All Exon Kits Version 2 Agilent, the Human All Exon V4+UTRs – 70 Mb Agilent, the Human All Exon Agilent, the Roche V.3, Medexome or the Twist Refseq kit, followed by a massively parallel sequencing on the HiSEQ 2000, or the NextSeq500 or NovaSeq system (Illumina). The mean coverage was 104.9X (range 51.1-225.8X) and 25, 30, 50-fold mean sequencing depth was achieved across 84.6% (range 79.5-90.4%), 86% (range 58.4-95.5%) and 81.4% (range 57-97.3%) of targeted regions, respectively. *DAGLB* specific coverage was analyzed using an in house script from the DAC core facility, based on transcript ENST00000297056 for exon and 5' and 3'UnTranslated Regions (UTR) localization. The mean coverage was 67.5X (range 12.9-209.2X), an example of the *DAGLB* coverage is shown on Supplementary Figure 1. Read alignment and variant calling were done using an in-house pipeline. Briefly, FastQC was used to check the quality of the reads and low-quality reads were removed using Trimmomatic. Sequencing data were then aligned to the human reference genome hg19 using the bwa suite[36] and variant calling was performed using GATK HaplotypeCaller[5] or Dragen (Illumina).

For 362 index cases, 25 additional patients and 38 unaffected relatives we were able to detect CNVs based on WES data. Briefly, the DRAGENTM DNA pipeline v3.8.4 (Illumina) was used to align the reads to the human hg19 reference genome, mark the PCR duplicates and perform the calling of the Copy Number Variants using the panel of normals approach. Each sample's depth of coverage being first corrected for the GC bias and then normalized against the depth of all the unrelated samples in the same sequencing batch. Only the events passing the default filters were considered for analysis and annotated with AnnotSV v3.1.1[6]. For the other individuals of the cohort (n=344), it was not possible due to the fact that too few individuals were sequenced together in the same batch.

Variants in *DAGLB* were extracted using the graphical interface developed by the DAC core facility (https://quby.icm-institute.org/app/dejavu) and annotated with a in house R script. We filtered all homozygous or double heterozygous variants in *DAGLB* that possibly affected the cDNA or localized in the splice site region (-8 +11bp from exon/intron junction) and with a

MAF<1% in the gnomAD public database. All variants identified in our cohort, including heterozygous variants, are listed in Supplementary Table 5.

For patient SPD-2137-1, WES was analyzed using VaraFT[7] softaware. We filtered all homozygous variants that possibly affected the cDNA or localized in the splice site region (-8 +11bp from exon/intron junction), with a MAF<1% in the gnomAD public database and and localized in region of loss of heterozygosity (Supplementary Table 6).

**Detection of run of homozygosity**

For patient SPD-2137_1 homozygosity mapping was performed using Automap[8] from the VCF file of WES data. Region of homozygosity are available on the Supplementary Table 3.

**Variant segregation**

For individuals from family SPD-2137 and individuals listed in Supplementary Table 5, the presence of the variant was checked on the bam file using IGV software (https://software.broadinstitute.org/software/igv/).

**3D structural modelling of DAGLB and predicted impact of the mutation**

DAGLB 3D protein structure was downloaded from Uniprot using the structure predict by AlphaFold[9] (Q8NCG7). Then, the impact of the mutation was predict using DynaMut2[10] and PremPS[11]

**Supplementary Table 1**. Pathogenicity score and prediction for the chr7:g.6464453G>A variant of DAGLB obtained with dbNSFP4.3a[12–14]

| | |
|---|---|
| gene_id | DAGLB |
| genomic | chr7:g.6464453G>A |
| hgvs_c | c.1070C>T |
| hgvs_p | p.Pro357Leu |
| Eigen-raw_coding | 0.84 |
| Eigen-phred_coding | 9.66 |
| SIFT4G_score | 0.00 |
| SIFT4G_pred | Damaging |
| Polyphen2_HDIV_score | 1.0 |
| Polyphen2_HDIV_pred | probably Damaging |
| MutationTaster_score | 1.00 |
| MutationTaster_pred | Disease_causing |
| MutationAssessor_score | 3.175 |
| MutationAssessor_pred | Medium |
| LRT_score | 0.00 |
| LRT_pred | Deleterious |
| CADD_raw | 3.97 |
| CADD_phred | 26.8 |
| MetaLR_score | 0.42 |
| MetaLR_pred | Tolerated |
| M-CAP_score | 0.08 |
| M-CAP_pred | Damaging |
| MutPred_AAchange | P357L |
| MutPred_score | 0.70 |
| MutPred_Top5features | Loss of ubiquitination at K352 (P = 0.0787) |
| REVEL_score | 0.66 |
| PROVEAN_score | -8.68 |
| PROVEAN_pred | Damaging |
| VEST4_score | 0.98 |
| MetaSVM_score | -0.10 |
| MetaSVM_pred | Tolerated |
| PrimateAI_score | 0.72 |
| PrimateAI_pred | Tolerated |
| MVP_score | 0.87 |
| DEOGEN2_score | 0.54 |
| DEOGEN2_pred | Damaging |
| ClinPred_score | 0.98 |
| ClinPred_pred | Deleterious |

.Predicted deleterious score are indicated in Yellow, predicted tolerated score are indicated in green.

**Supplementary Table 2.** Coding or splice site variants, absent from GnomAD in the homozygous state, identified in the WES analyses of patient SPD-2137-1

| Nature of the variants | hg19 genomic position | gene_id | hgvs_c | hgvs_p | Htz GnomAD | Hmz GnomAD | Chr GnomAD |
|---|---|---|---|---|---|---|---|
| missense_variant | chr1:g.1961503C>T | GABRD | c.1141C>T | p.Arg381Cys | 5 | 0 | 276182 |
| synonymous_variant | chr6:g.33643510C>T | ITPR3 | c.3159C>T | p.Arg1053Arg | 103 | 0 | 280076 |
| missense_variant | chr6:g.33996029C>A | GRM4 | c.2557G>T | p.Val853Leu | 15 | 0 | 251320 |
| missense_variant | chr6:g.42072819A>G | C6orf132 | c.2831T>C | p.Val944Ala | 21 | 0 | 173072 |
| missense_variant | chr6:g.43251078C>T | TTBK1 | c.2600C>T | p.Pro867Leu | 23 | 0 | 237208 |
| synonymous_variant | chr7:g.4839063C>T | RADIL | c.3174G>A | p.Ala1058Ala | 110 | 0 | 280618 |
| missense_variant | chr7:g.6464453G>A | DAGLB | c.1070C>T | p.Pro357Leu | 5 | 0 | 282848 |
| missense_variant | chr7:g.149489813C>T | SSPO | c.5869C>T | p.Arg1957Cys | 227 | 0 | 213992 |
| missense_variant | chr16:g.10729739T>G | TEKT5 | c.1123A>C | p.Met375Leu | 49 | 0 | 282730 |
| stop_gained&splice_region_variant | chr16:g.57448995C>T | CCL17 | c.73C>T | p.Arg25* | 8 | 0 | 282676 |
| missense_variant&splice_region_variant | chr16:g.57998447G>C | CNGB1 | c.161C>G | p.Pro54Arg | 1 | 0 | 249466 |
| missense_variant | chr16:g.58043902G>A | USB1 | c.335G>A | p.Arg112Gln | 37 | 0 | 282838 |
| missense_variant | chr19:g.48305050C>A | TPRX1 | c.1218G>T | p.Arg406Ser | 28 | 0 | 281736 |
| missense_variant | chr19:g.49565052C>T | NTF4 | c.203G>A | p.Arg68Gln | 197 | 0 | 187424 |
| missense_variant | chr19:g.49655273C>T | HRC | c.2014G>A | p.Val672Ile | 483 | 0 | 282296 |
| missense_variant | chr19:g.50755969A>G | MYH14 | c.1904A>G | p.His635Arg | NA | NA | NA |
| missense_variant | chr19:g.56538791G>A | NLRP5 | c.1192G>A | p.Val398Ile | 11 | 0 | 279368 |
| missense_variant | chr19:g.56599855G>C | ZNF787 | c.686C>G | p.Ala229Gly | 0 | 0 | 22454 |
| missense_variant | chr19:g.57641219T>G | USP29 | c.1176T>G | p.Asn392Lys | 9 | 0 | 250986 |
| missense_variant | chr19:g.57642033G>A | USP29 | c.1990G>A | p.Glu664Lys | 9 | 0 | 251082 |
| frameshift | chr6:g.27861345_27861346delAG | H2BC17 | c.109_110del | p.(Ser37Leufs*32) | 4 | 0 | 251452 |

This table was created with the R package myvariant 1.20.0[15].

**Supplementary Table 3.** Clinical data of the patient SPD-2137-1 and the patients of the lead article. Adapted from Liu et al, 2022[16]

| | Liu *et al*, 2022 | | | | | | Our case |
|---|---|---|---|---|---|---|---|
| | Family 1<br>II-3 | Family 1<br>II-4 | Family 2<br>II-4 | Family 3<br>II-1 | Family 3<br>II-2 | Family 4<br>II-1 | SPD-2137-1 |
| Mutation | c.1821-2A>G | c.1821-2A>G | c.1088A>G<br>p.Asp363Gly | g.chr7:6,486,383-6,489,136 del | g.chr7:6,486,383-6,489,136 del | c.469dupC<br>p.Leu158Serfs*17 | c.1070C>T<br>p.Pro357Leu |
| Age at onset (years) | <40 | <40 | <40 | <40 | <40 | <40 | 30 |
| Age at examination (year) | NA | NA | NA | NA | NA | NA | 43 |
| Symptoms at onset | Resting tremor | Resting tremor | Bradykinesia | Resting tremor | Resting tremor | Bradykinesia | Bradykinesia |
| Asymmetry at onset | + | + | + | + | + | + | + |
| Hoehn-Yahr stage (Off/On) | IV/II | IV/II | IV/II | III/II | V/III | III/II | IV/III |
| **Motor symptom** | | | | | | | |
|     Bradykinesia | + | + | + | + | + | + | + |
|     Resting tremor | + | + | + | + | + | + | + |
|     Rigidity | + | + | + | + | + | + | + |
|     Postural instability | + | + | + | + | + | + | + |
|     Hypomimia | + | + | + | + | + | + | + |
|     UPDRS III (Off/On) | 56/37 | 74/40 | 76/33 | 52/30 | 76/58 | 50/28 | 78/45 |
| **Nonmotor symptom** | | | | | | | |
|     Depression | + | + | + | + | + | + | + |
|     Urinary urgency | + | + | + | - | + | + | + |
|     Constipation | - | + | + | - | + | + | + |
|     Cognitive decline | - | - | - | + | - | - | - |
|     Hallucination | + | - | - | - | - | - | - |
|     Sleep disturbance | + | + | + | + | + | + | + |
|     Freezing gait | + | + | + | + | + | + | + |
|     RBD | - | + | + | + | - | - | - |
| Response to levodopa | + | + | + | + | + | + | + |
| Complications with treatment | | | | | | | |

| | | | | | | | | |
|---|---|---|---|---|---|---|---|---|
| Wearing off | + | + | + | + | + | + | + | + |
| On-off phenomenon | + | + | + | + | + | + | + | + |
| Dyskinesia | + | + | + | - | - | + | + | + |
| Surgical therapies | NA | NA | +* | NA | +# | NA | NA | NA |
| **Brain MRI** | - | - | - | - | - | - | - | Normal |
| **Brain 11C-CFT PET** | NA | NA | Abnormal* | NA | NA | NA | NA | NA |

+ = Present; - = Absent; NA=Not performed

\* Deep brain stimulation

\# Posteroventral pallidotomy

\*Severe striatal uptake deficit, particularly at putamen level, as seen in Parkinson's disease

MRI=magnetic resonance imaging; PET=positron emission tomography; CFT=C-2β-carbomethoxy-3β-(4-fluorophenyl) tropane

**Supplementary Table 4.** Known genes or genes related to Parkinson's disease/Parkinsonism excluded from our cohort

| Gene | Transmission | Designation | NM number |
|---|---|---|---|
| *SNCA* | AD | PARK-*SNCA* | NM_000345.3 |
| *PRKN* | AR | PARK-*Parkin* | NM_004562.2 |
| *PINK1* | AR | PARK-*PINK1* | NM_032409.2 |
| *PARK7* | AR | PARK-*DJ1* | NM_007262.4 |
| *LRRK2* | AD | PARK-*LRRK2* | NM_198578.3 |
| *ATP13A2* | AR | MxMD-*ATP13A2* | NM_022089.2 |
| *UCHL1* | AD | HSP/ATX-*UCHL1* | NM_004181.4 |
| *GIGYF2* | AD | Risk Factor | NM_001103147.1 |
| *HTR2A* | AD | Risk Factor | NM_000621.4 |
| *PLA2G6* | AR | MxMD-*PLA2G6* | NM_003560.2 |
| *FBXO7* | AR | PARK-*FBXO7* | NM_012179.3 |
| *VPS35* | AD | PARK-*VPS35* | NM_018206.5 |
| *EIF4G1* | AD | Risk Factor | NM_182917.4 |
| *DNAJC6* | AR | PARK-*DNAJC6* | NM_001256864.1 |
| *SYNJ1* | AR | PARK-*SYNJ1* | NM_003895.3 |
| *SLC6A3* | AR | DYT/PARK-*SLC6A3* | NM_001044.4 |
| *VPS13C* | AR | PARK-*VPS13C* | NM_020821.2 |
| *TMEM230* | AD | Risk Factor | NM_001009923.1 |
| *RIC3* | AD | | NM_024557.4 |
| *CHCHD2* | AD | PARK-*CHCHD2* | NM_001320327.1 |
| *DNAJC13* | AD | Risk Factor | NM_001329126.1 |
| *GBA1* | Mu/AD | PARK-*GBA* | NM_001005741.2 |
| *ADH1C* | Mu/AD | | NM_000669.4 |
| *TBP* | Mu/AD | SCA-*TBP* | NM_003194.4 |
| *ATXN2* | Mu/AD | SCA-*ATXN2* | NM_002973.3 |
| *MAPT* | Mu/AD | | NM_001123066.3 |
| *GCH1* | AD/AR | DYT/PARK-*GCH1* | NM_001024024.1 |
| *DCTN1* | AD/AR | PARK-*DCTN1* | NM_004082.4 |
| *PANK2* | AR | DYT-*PANK2*-(NBIA) | NM_153638.2 |
| *POLG* | AR/AD | MxMD-*POLG* | NM_002693.2 |
| *SPG11* | AR | SPG-*KIAA1840* | NM_025137.3 |
| *TH* | AR | DYT/PARK-*TH* | NM_199292.2 |
| *ADORA1* | AR | | NM_000674.2 |
| *RAB39B* | XLR | PARK-*RAB39B* | NM_171998.3 |
| *TNK2* | AD | | NM_001010938.1 |
| *PODXL* | AD | | NM_001018111.2 |
| *TNR* | AD | | NM_003285.2 |
| *PTRHD1* | AR | | NM_001013663.2 |

AD, autosomal dominant; AR, autosomal recessive, Mu, Multifactorial according to OMIM (https://www.omim.org). Designation were obtained using recommandation[17–20].

**Supplementary Table 5.** Rare *DAGLB* variants identified in our cohort of 716 patients with PD from 683 families

| genomic | hgvs_c | hgvs_p | Ind | Zygosity | GnomAD Htz/Hmz/Chr | Pathogenicity | ACMG classifictaion | Index cases | Sex | Statut | Ethnicity | AAO | Inheritance | Consanguinity |
|---|---|---|---|---|---|---|---|---|---|---|---|---|---|---|
| chr7:g.6449476C>T | c.2011G>A | p.Val671Met | SPD-755-16 | htz | 12/0/281876 | 3/17 | VUS | Yes | Male | Affected | CAU | 45 | Isolated case | No |
| chr7:g.6449816G>C | c.1765C>G | p.Pro589Ala | 6619NG003901 | htz | 44/0/260808 | 3/18 | VUS | Yes | Male | Affected | NA | 32 | Isolated case | No |
| chr7:g.6449930C>T | c.1651G>A | p.Asp551Asn | G1801 | htz | 22/0/282008 | 0/17 | VUS | Yes | Male | Affected | CAU | 45 | Isolated case | No |
| chr7:g.6449947G>T | c.1634C>A | p.Thr545Lys | SPD-363-1 | htz | 271/0/282344 | 8/17 | VUS | Yes | Male | Affected | CAU | 41 | Isolated case | No |
| chr7:g.6461357C>A | c.1218+1G>T | p.( ?) | 16NG001469 | htz | Absent | 3/3 | VUS | Yes | Male | Affected | NA | 20 | Isolated case | Yes |
| chr7:g.6464453G>A | c.1070C>T | p.Pro357Leu | SPD-2137-2 | htz | 5/0/282848 | 13/18 | VUS | No | Male | Not_Affected | NoAf |  |  | No |
| chr7:g.6464453G>A | c.1070C>T | p.Pro357Leu | SPD-2137-1 | hmz | 5/0/282848 | 13/18 | VUS | Yes | Male | Affected | NoAf | 30 | Isolated case | Yes |
| chr7:g.6472577A>G | c.692T>C | p.Val231Ala | PD-TU-28 | htz | 42/0/281998 | 12/17 | VUS | Yes | Female | Affected | NoAf | 40 | Isolated case | No |
| chr7:g.6472577A>G | c.692T>C | p.Val231Ala | SPD-492-11 | htz | 42/0/281998 | 12/17 | VUS | Yes | Male | Affected | NoAf | 19 | Isolated case | Yes |
| chr7:g.6472577A>G | c.692T>C | p.Val231Ala | SPD-1547-1 | htz | 42/0/281998 | 12/17 | VUS | Yes | Male | Affected | NA | 34 | Isolated case | Yes |
| chr7:g.6472577A>G | c.692T>C | p.Val231Ala | SPD-2166-7 | htz | 42/0/281998 | 12/17 | VUS | Yes | Male | Affected | NoAf | 26 | Isolated case | Yes |
| chr7:g.6472577A>G | c.692T>C | p.Val231Ala | SPD-2158-6 | htz | 42/0/281998 | 12/17 | VUS | Yes | Male | Affected | NoAf | 45 | Isolated case | No |
| chr7:g.6474561del | c.510del | p.(Ser171Alafs*105) | SPD-1597-1 | htz | Absent | NA | VUS | Yes | Female | Affected | NA | 27 | Isolated case | ND |
| chr7:g.6476030T>C | c.382A>G | p.Thr128Ala | 16NG003029 | htz | Absent | 1/18 | VUS | Yes | Male | Affected | NA | 40 | Isolated case | No |
| chr7:g.6476140C>G | c.272G>C | p.Arg91Pro | 17NG001059 | htz | Absent | 16/18 | VUS | Yes | Male | Affected | NA | 49 | Isolated case | No |
| chr7:g.6476161G>C | c.251C>G | p.Thr84Arg | 6619NG003284 | htz | 712/3/282162 | 9/17 | VUS | Yes | Female | Affected | NA | 23 | Isolated case | No |

This table was created with the R package myvariant 1.20.0[15]. ACMG[22] nomenclature was obtained using Franklin by genoox (https://franklin.genoox.com/clinical-db/home), VUS : Variant of unknown significance, CAU : Caucasian origin, NoAf : North African origin, NA : Information not available.

**Supplementary Table 6.** List of homozygosity regions for patient SPD-2137-1 detected with Automap[8]

| #Chr | Begin | End | Size(Mb) | Nb_variants | Percentage_homozygosity |
|---|---|---|---|---|---|
| chr1 | 69270 | 3342804 | 3.27 | 114 | 94.74 |
| chr1 | 193051685 | 199717218 | 6.67 | 43 | 93.02 |
| chr6 | 18264210 | 31238983 | 12.97 | 231 | 97.4 |
| chr6 | 31324144 | 32489672 | 1.17 | 113 | 93.81 |
| chr6 | 32551961 | 45320548 | 12.77 | 228 | 96.93 |
| chr7 * | 2946461 | 11075263 | 8.13 | 82 | 93.9 |
| chr7 | 149191568 | 150439500 | 1.25 | 33 | 90.91 |
| chr11 | 45230862 | 52516144 | 7.29 | 75 | 93.33 |
| chr12 | 50040811 | 51393116 | 1.35 | 34 | 88.24 |
| chr12 | 123087442 | 124246910 | 1.16 | 33 | 90.91 |
| chr13 | 103473497 | 113665848 | 10.19 | 96 | 96.87 |
| chr14 | 100772102 | 103799695 | 3.03 | 37 | 89.19 |
| chr16 | 5140541 | 15091589 | 9.95 | 122 | 98.36 |
| chr16 | 56504724 | 59757683 | 3.25 | 66 | 98.48 |
| chr18 | 4197503 | 11610332 | 7.41 | 55 | 98.18 |
| chr18 | 11610629 | 28681903 | 17.07 | 84 | 96.43 |
| chr19 | 45515345 | 52130488 | 6.62 | 288 | 99.31 |
| chr19 | 54849481 | 59080698 | 4.23 | 257 | 100 |
| chr20 | 59485627 | 62737568 | 3.25 | 134 | 100 |
| chr22 | 47132809 | 51183255 | 4.05 | 80 | 98.75 |

Default settings used:
## INFO: 125.08 Mb are in Homozygous Regions (autosomal chromosomes)
## AutoMap v1.0 used for analysis
## Variant filtering parameters used: DP=8, percaltlow=.25, percalthigh=.75, binomial=.000001, maxgap=10
## Other parameters used: window=7, windowthres=5, minsize=1, minvar=25, minperc=88, chrX=No, extend=1
* *DAGLB* localization

**Supplementary Figure 1.** Example of DAGLB coverage for one individuals of our cohort that shown that all exons are well covered. The uncover region is limited to the end of the 3'UTR.

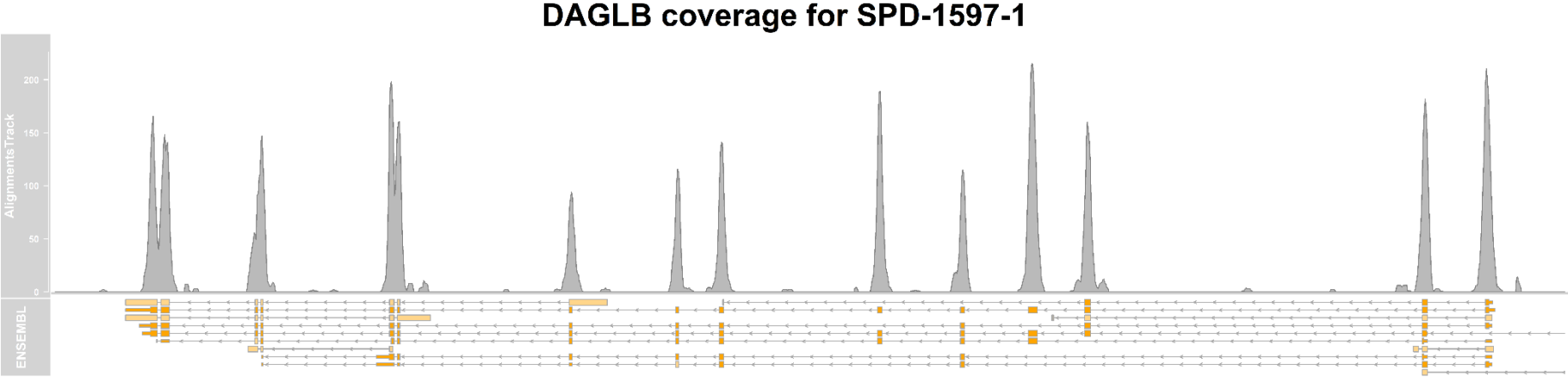

Figure obtained with the package R Gviz[21].

**Supplementary references**

1. Lesage S, Lunati A, Houot M, et al. Characterization of Recessive Parkinson Disease in a Large Multicenter Study. *Ann Neurol*. 2020;88(4):843-850. doi:10.1002/ana.25787

2. Lesage S, Houot M, Mangone G, et al. Genetic and Phenotypic Basis of Autosomal Dominant Parkinson's Disease in a Large Multi-Center Cohort. *Front Neurol*. 2020;11:682. doi:10.3389/fneur.2020.00682

3. Dolzhenko E, Deshpande V, Schlesinger F, et al. ExpansionHunter: a sequence-graph-based tool to analyze variation in short tandem repeat regions. *Bioinformatics*. 2019;35(22):4754-4756. doi:10.1093/bioinformatics/btz431

4. Casse F, Courtin T, Tesson C, et al. Detection of ATXN2 Expansions in an Exome Dataset: An Underdiagnosed Cause of Parkinsonism. *Mov Disord Clin Pract*. 2023;10(4):664-669. doi:10.1002/mdc3.13699

5. McKenna A, Hanna M, Banks E, et al. The Genome Analysis Toolkit: a MapReduce framework for analyzing next-generation DNA sequencing data. *Genome Res*. 2010;20(9):1297-1303. doi:10.1101/gr.107524.110

6. Geoffroy V, Herenger Y, Kress A, et al. AnnotSV: an integrated tool for structural variations annotation. *Bioinformatics*. 2018;34(20):3572-3574. doi:10.1093/bioinformatics/bty304

7. Desvignes JP, Bartoli M, Delague V, et al. VarAFT: a variant annotation and filtration system for human next generation sequencing data. *Nucleic Acids Research*. 2018;46(W1):W545-W553. doi:10.1093/nar/gky471

8. Quinodoz M, Peter VG, Bedoni N, et al. AutoMap is a high performance homozygosity mapping tool using next-generation sequencing data. *Nat Commun*. 2021;12:518. doi:10.1038/s41467-020-20584-4

9. Jumper J, Evans R, Pritzel A, et al. Highly accurate protein structure prediction with AlphaFold. *Nature*. 2021;596(7873):583-589. doi:10.1038/s41586-021-03819-2

10. Rodrigues CHM, Pires DEV, Ascher DB. DynaMut2: Assessing changes in stability and flexibility upon single and multiple point missense mutations. *Protein Sci*. 2021;30(1):60-69. doi:10.1002/pro.3942

11. PremPS: Predicting the impact of missense mutations on protein stability - PubMed. Accessed February 2, 2023. https://pubmed-ncbi-nlm-nih-gov.proxy.insermbiblio.inist.fr/33378330/

12. Liu X, Jian X, Boerwinkle E. dbNSFP: a lightweight database of human nonsynonymous SNPs and their functional predictions. *Hum Mutat*. 2011;32(8):894-899. doi:10.1002/humu.21517

13. Liu X, Li C, Mou C, Dong Y, Tu Y. dbNSFP v4: a comprehensive database of transcript-specific functional predictions and annotations for human nonsynonymous and splice-site SNVs. *Genome Med*. 2020;12(1):103. doi:10.1186/s13073-020-00803-9

14. Dong C, Wei P, Jian X, et al. Comparison and integration of deleteriousness prediction methods for nonsynonymous SNVs in whole exome sequencing studies. *Human Molecular Genetics*. 2015;24(8):2125-2137. doi:10.1093/hmg/ddu733

15. Adam M. myvariant: Accesses MyVariant.info variant query and annotation services. R package version 1.20.0.

16. Liu Z, Yang N, Dong J, et al. Deficiency in endocannabinoid synthase DAGLB contributes to early onset Parkinsonism and murine nigral dopaminergic neuron dysfunction. *Nat Commun*. 2022;13:3490. doi:10.1038/s41467-022-31168-9

17. van der Veen S, Zutt R, Klein C, et al. Nomenclature of Genetically Determined Myoclonus Syndromes: Recommendations of the International Parkinson and Movement Disorder Society Task Force. *Movement Disorders*. 2019;34(11):1602-1613. doi:10.1002/mds.27828

18. Lange LM, Gonzalez-Latapi P, Rajalingam R, et al. Nomenclature of Genetic Movement Disorders: Recommendations of the International Parkinson and Movement Disorder Society Task Force – An Update. *Movement Disorders*. 2022;37(5):905-935. doi:10.1002/mds.28982

19. Marras C, Lang A, van de Warrenburg BP, et al. Nomenclature of genetic movement disorders: Recommendations of the international Parkinson and movement disorder society task force. *Movement Disorders*. 2016;31(4):436-457. doi:10.1002/mds.26527

20. Rossi M, Anheim M, Durr A, et al. The genetic nomenclature of recessive cerebellar ataxias. *Movement Disorders*. 2018;33(7):1056-1076. doi:10.1002/mds.27415

21. Hahne F, Ivanek R. Visualizing Genomic Data Using Gviz and Bioconductor. *Methods Mol Biol*. 2016;1418:335-351. doi:10.1007/978-1-4939-3578-9_16

22. Richards S, Aziz N, Bale S, et al. Standards and guidelines for the interpretation of sequence variants: a joint consensus recommendation of the American College of Medical Genetics and Genomics and the Association for Molecular Pathology. *Genet Med*. 2015;17(5):405-424. doi:10.1038/gim.2015.30